# Strong near-field light-matter interaction in plasmon-resonant tip-enhanced Raman scattering in indium nitride

Emanuele Poliani[1,2,*], Daniel Seidlitz[1], Maximilian Ries[1], Soo J. Choi[3], James S. Speck[3], Axel Hoffmann[1], and Markus R. Wagner[1]

[1] Institute of Solid State Physics, Technische Universität Berlin, 10623 Berlin, Germany
[2] School of Chemistry, Manchester University, Manchester M139PL, United Kingdom
[3] Materials Department, University of California, Santa Barbara, California 93106-5050, USA

* corresponding author: emanuele.poliani@gmail.com

**Abstract**

We report a detailed study of the strong near-field Raman scattering enhancement which takes place in tip-enhanced Raman scattering (TERS) in indium nitride. In addition to the well-known first-order optical phonons of indium nitride, near-field Raman modes, not detectable in the far-field, appear when approaching the plasmonic probe. The frequencies of these modes coincide with calculated energies of second order combinational modes consisting of optical zone center phonons and acoustic phonons at the edge of the Brillouin zone. The appearance of strong combinational modes suggests that TERS in indium nitride represents a special case of Raman scattering in which a resonance condition on the nanometer scale is achieved between the localized surface plasmons (LSPs) and surface plasmon polaritons (SPPs) of the probe with the surface charge oscillation of the material. We suggest that the surface charge accumulation (SCA) in InN, which can render the surface a degenerate semiconductor, is the dominating reason for the unusually large enhancement of the TERS signal as compared to other inorganic semiconductors. Thus, the plasmon-resonant TERS (PR-TERS) process in InN makes this technique an excellent tool for defect characterization of indium-rich semiconductor heterostructures and nanostructures with high carrier concentrations.

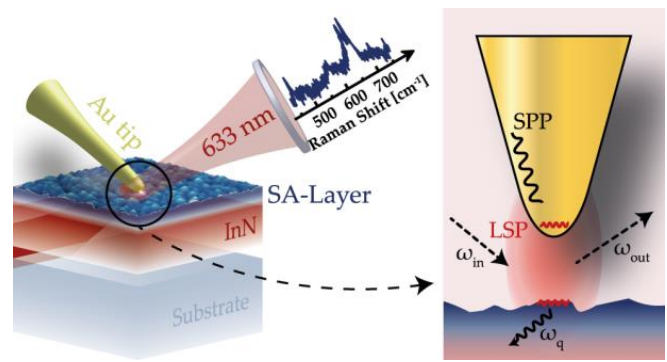

TOC Graphic

**Introduction**

During the last two decades, group-III-nitride semiconductor hetero- and nanostructures have established a dominant role as versatile materials for solid-state light-emitting devices. The large tunability of the direct band gap in the ternary alloys InGaN and AlGaN combined with the existence of shallow donor and acceptor dopants makes these ternary material systems the best available choices for optoelectronic applications ranging from the ultraviolet to the green spectral region.[1] The bottleneck, which prevents nitride-based devices to be available as longer wavelengths emitters is the systematic reduction of the internal quantum efficiency (IQE) with increasing emission wavelength (green gap), as well as with increased drive current (droop).[2] Among the reasons for these limitations are hole-localization, alloy composition fluctuations, and In clustering, resulting in an increase of non-radiative recombination mechanisms.[2,3] These effects become more pronounced for alloys with high In content and contribute to the limitations of long-wavelength light-emitters based on the InGaN material system. Even in the case of pure InN, nanoclusters of metallic In may be present, which modify the electronic and optical properties of the material.[4,5,6]

Raman spectroscopy is recognized as an essential tool for the characterization of strain, compositional fluctuations and defects in semiconductors, however, it remains a diffraction-limited technique, which cannot access subwavelength information, i.e. it is not suitable for characterization on the nanometer scale. Optical near-field techniques are powerful solutions that circumvent these limitations and enable the characterization of nanostructures with few nanometers of spatial resolution. Tip-enhanced Raman spectroscopy (TERS) exploits the plasmonic and polaritonic properties of a metallic tip in order to enhance and localize the electromagnetic field which interacts with the sample.[7] While the majority of TERS studies focus on the vibrational properties of molecules and clusters,[8,9,10] only few works exist that address near-field Raman scattering in inorganic semiconductor and their nanostructures. These include Si,[11,12] Ge and SiGe,[13,14] GaN,[7,15] $BaTiO_3$,[16] GaAs,[17] as well as nanowires of InP,[18] GaN,[19] and Ge.[20] In the case of the III-nitride material system, we have previously reported the occurrence of InN clusters in InGaN nanostructures using TERS and derived detailed information on the strain distribution, chemical composition, polymorphism, and charge accumulation with a spatial resolution below 35 nm.[21]

In this work, we demonstrate the potential of TERS for the nanoscale characterization of III-nitrides by studying the plasmonic and polaritonic interaction between InN and a gold tip. We show how the strong plasmon-resonant interaction between the plasmonic probe and the surface charges in the surface accumulation (SA) layer of InN leads to a tremendous increase of the Raman intensity of combinational optical and acoustic phonon modes. The plasmonic mediation strongly modifies the Raman scattering probability which is theoretically described by including plasmonic energy levels into the Hamiltonian.[7,22] We argue that the resonance condition achieved between the InN surface plasmons, the gold tip plasmons, and the laser excitation increases the electron-phonon coupling as the resonance between laser excitation and electronic structure does in canonical resonant Raman scattering. As expected for a resonant process, the Raman selection rules are relaxed and the probability for 2nd-order scattering is increased. The results are discussed considering the resonance between laser excitation, probe plasmons and surface charge carriers in the samples.

**Methods**

Epitaxial InN layers were grown by migration-enhanced plasma-assisted metal organic vapor deposition (MEPA-MOCVD) and plasma-assisted molecular beam epitaxy (PA-MBE). The MEPA sample is composed of a 90 nm thick InN heteroepitaxial film grown on a c-plane sapphire substrate.[23] It contains a bulk free carrier density of around $0.8 - 1 \times 10^{20}$ cm$^{-3}$ as determined by Raman and FTIR measurements.[23] The MBE sample consists of a 2 µm thick InN film grown on a 100 nm buffer layer of C-doped GaN. The buffer layer was deposited on a 3.5 µm thick Fe-doped GaN template on c-plane sapphire commercially available from Lumilog. It contains a bulk free carrier density of around $5.2 \times 10^{17}$ cm$^{-3}$ as determined by Hall effect measurements. Residual indium was etched using HCl after unloading.[24]

TERS measurements are conducted using a commercial Raman spectrometer (Horiba-Jobin Yvon LabRam HR-800) coupled with an atomic force microscope (Park XE-100 AFM). An 80x long-distance microscope objective, tilted by 60° degrees with respect to the z-axis, illuminates the bulk gold tip produced by electrochemical etching of a 50 µm gold wire with the light from a 633 nm He-Ne laser. The near-field Raman signal generated in the tip-sample junction is collected by the same objective. A detailed description of the far-field scattering geometry and the near-field signal acquisition is reported elsewhere.[7]

**Results and discussion**

Fig. 1a and Fig. 1b display the Raman spectra of the MEPA sample and the MBE sample, respectively. The black curves represent the far-field spectra acquired in the z(-,-)z backscattering geometry. Both samples show the $E_2$(high) and $A_1$(LO) Raman modes of InN in the displayed spectral range between 400 and 750 cm$^{-1}$.[24] A direct comparison between the samples reveals a smaller FWHM of the $E_2$(high) mode in the MBE sample, indicating a superior "bulk" crystalline quality as compared to the MEPA sample. The blue curves show the far-field spectra acquired with the 60° tilted TERS objective and retracted gold tip, which are displayed as reference for the near-field spectra. By approaching the tip to the surface of the sample in this geometry, the TERS spectra are obtained (red curves in Fig. 1). Both samples exhibit a drastic change of their Raman spectra expressed by variations of mode intensities and the appearance of new peaks in the near-field. This observation is in agreement with previously reported TERS spectra of InN clusters embedded in InGaN quantum wells.[21] The reason for the pronounced change of the Raman spectra lies within the increased surface sensitivity (probing-depth approximately 5-10 nm) and greatly reduced scattering volume (nanometer scale)[25] as well as the breakdown of classical Raman selection rules in near-field Raman scattering.[7]

In the following paragraphs, we discuss the near-field Raman spectra of the different InN samples in detail. The $A_1$(TO) phonon is strongly enhanced in both near-field spectra, disregarding the far-field selection rules. Based on the smaller FWHM of the $A_1$(TO) mode we conclude that the crystal quality in the surface region that gives rise to the TERS signal is higher in the MEPA sample than in the MBE sample. This is in alignment with the surface roughness measured by AFM. The MEPA sample exhibits a rather smooth surface (mean surface roughness $R_{RMS}$ = 3.7 nm), whereas the MBE sample shows domains of different heights possibly related to the post growth chemical etching ($R_{RMS}$ = 11.8 nm). A mode close to the frequency of the $E_1$(TO) phonon is observed in the MEPA sample, but is absent in the MBE sample. On the contrary, a mode close to the reported position of the $E_2$(high) is observed in the MBE sample but absent in the MEPA sample.[24] Based on the pronounced spatial variation of this mode, the mode is not assigned to a lattice vibration of InN but might be related to local surface contaminations. Furthermore, a broad Raman mode between 500 cm$^{-1}$ and 600 cm$^{-1}$ is visible in both spectra but at different frequencies. Thakur *et al.* observed a similar feature in the far-field Raman spectra of InN and assigned it to the $B_1$(high) phonon.[26,27] The authors reported experiments over a set of samples with different free carrier concentrations ranging from $3.8 \times 10^{20}$ to $1.6 \times 10^{21}$ cm$^{-3}$ and they observed a strong interaction between free charge carriers and optical modes, even influencing the spectral characteristics.. Madapu et al. also ascribed a mode in this range to the $B_1$(high) phonon and highlighted the sensitivity towards resonant conditions based on the charge carrier concentrations in different samples.[28] Alarcon-Lladó et al. investigated the influence of surface charges in InN on the behavior of the LO modes.[29] Besides the observation of the surface charge concentration dependent shift of the LO-modes, they reported the occurrence of the $B_1$(high) mode as well and highlighted the importance of charge carriers. In the present case, the observed peaks in this spectral region could be strongly influenced by the surface charge accumulation; however, the available data is insufficient for an unambiguous attribution of the phononic origin of these bands.

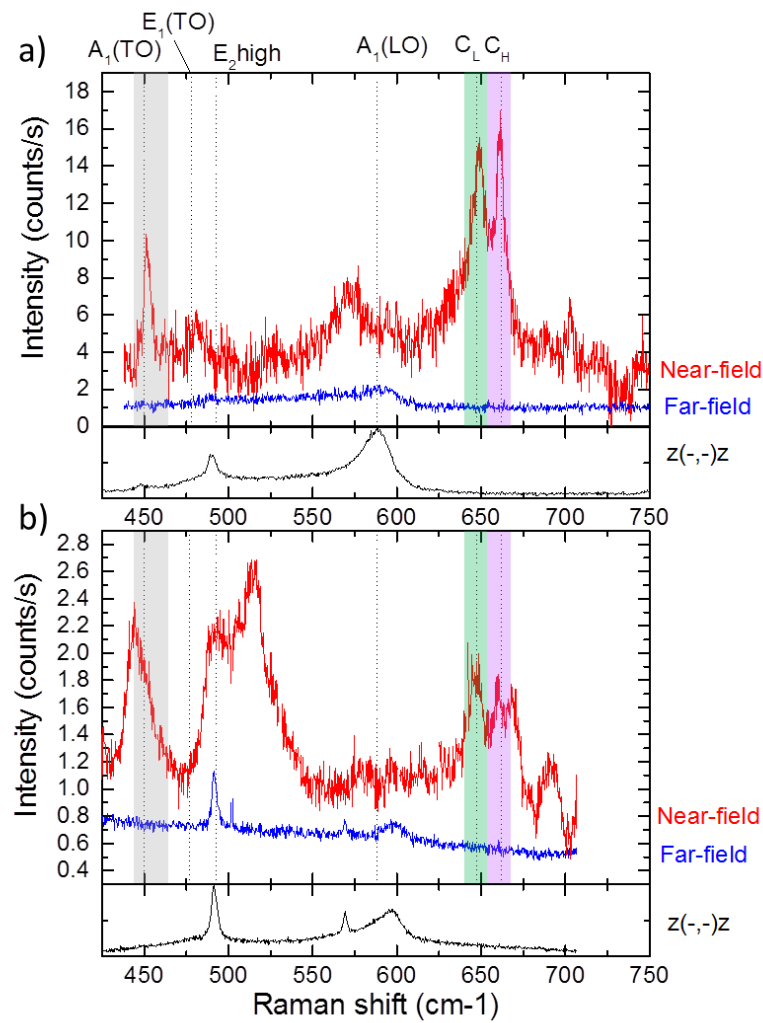

**Fig. 1** Backscattering z(-,-)z (black curve), far-field (blue curve) and near-field (red curve) Raman spectra of (a) 90 nm thick InN epitaxial layer grown by MEPA-MOCVD on sapphire (MEPA sample) and (b) 2 μm thick InN grown by PA-MBE on C-doped GaN (MBE sample). The near-field (red) curves represent the TERS spectra acquired with the gold tip approached. The shaded areas indicate the spectral region of the $A_1$(TO) (gray), $C_L$ (green), and $C_H$ (purple) Raman modes.

Apart from the first order Raman modes, two previously unreported peaks, labeled $C_L$ and $C_H$, appear in the near-field Raman spectra of both InN samples with average spectral positions of 648 cm$^{-1}$ and 664 cm$^{-1}$ (see

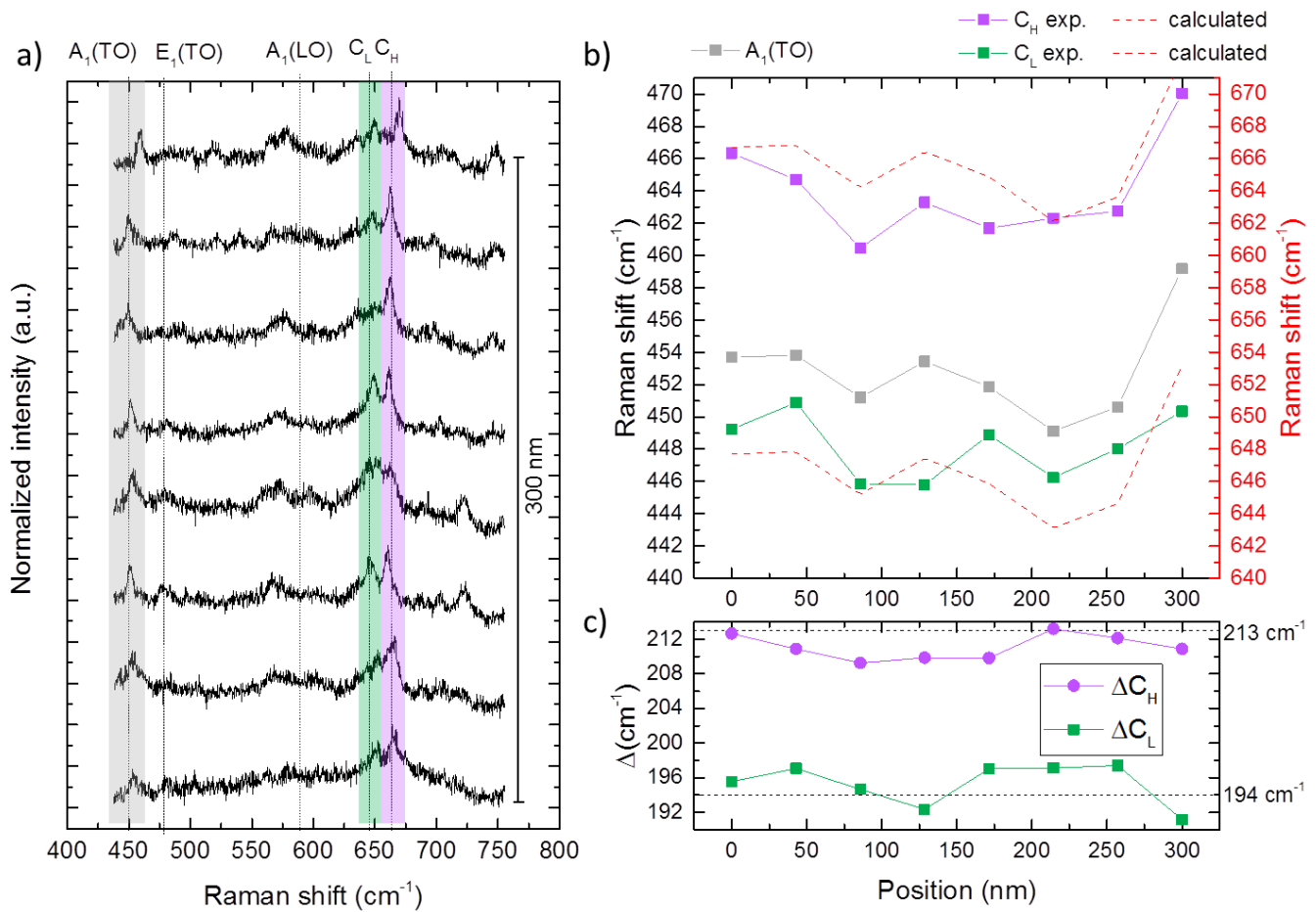

**Fig. 2** (a) TERS line scan of the MEPA sample over a distance of 300 nm with 8 steps resulting in a point-to-point distance of 43 nm. The $A_1$(TO) mode varies around its nominal position. The shaded areas indicate the spectral region of the $A_1$(TO) (gray), $C_L$ (green), and $C_H$ (purple) Raman modes. (b) Measured phonon frequencies of the $A_1$(TO), $C_L$, and $C_H$ Raman modes at the different positions along the line scan. The dashed (red) lines indicate the calculated values for the corresponding two phonon combinational modes as described in the text. (c) Comparison of the experimental values $C_{H,L}$ - $A_1$(TO) (green and purple dots) with the calculated values of $M_L$ = 194 cm$^{-1}$ and $M_H$ = 213 cm$^{-1}$ (gray dotted lines).

Fig. 1 and Tab. 1). These modes deserve particular attention as no Raman modes of InN are expected at these frequencies, neither based on calculation of the phonon dispersion, nor based on other far-field experimental works.[24,30,31] However, the two-phonon density of states (2PDOS) exhibits pronounced features in this frequency range, which suggests a two phonon scattering process.[24] In order to understand the origin of these two Raman modes, we performed a TERS surface line scan on the MEPA sample with a point-to-point distance of 43 nm, as shown in Fig. 2. Over most of the distance of the line scan, the $A_1$(TO) phonon energy scatters equally around its nominal value and towards the end shifts to higher wavenumbers. These fluctuations of the $A_1$(TO) phonon energy are likely caused by local variations of strain and charge accumulation at the surface.[24,32] Interestingly, the Raman modes $C_L$ and $C_H$ seem to follow the same behavior. In order to explain this correlation, we compare the position and energy spacing of these modes.

The frequencies of the phonons $A_1$(TO), $C_L$, and $C_H$ as a function of the spatial coordinates are shown in Fig. 2b as obtained by peak fitting. Subtraction of the experimentally obtained values of the $A_1$(TO) mode from those of the $C_L$ and $C_H$ modes gives average values of 195 cm$^{-1}$ and 211 cm$^{-1}$, respectively. In this frequency range, a Raman mode around 200 cm$^{-1}$ was reported in the literature which was attributed either to the $B_1$(low) silent mode[33] or to overtones of transverse acoustic (TA) phonons near the symmetry points K or M of the Brillouin zone.[34] Comparing our experimental data with the phonon dispersion calculations reported in Ref.[31], Ref.[34], and Ref.[35], we suggest that the modes $C_L$ and $C_H$ are combinational modes composed of the $A_1$(TO) phonon with M point phonons originating from the longitudinal acoustic (LA) branch. We can therefore calculate the expected values for the combinational modes as the experimental value of the $A_1$(TO) phonon plus the calculated values for the acoustic phonons:

$$C_H^{expected} = A_1(TO)^{exper.} + M_H^{calc.}$$

$$C_L^{expected} = A_1(TO)^{exper.} + M_L^{calc.}. \quad \text{(eq. 1)}$$

We use the strain free values of 194 cm$^{-1}$ for the $M_L^{calc.}$ phonon and 213 cm$^{-1}$ for the $M_H^{calc.}$ phonon from Ref.[31] The impact of local strain variations can be estimated by consideration of the reported pressure coefficients of phonon modes of InN in the literature. Calculations for the phonon dispersion of the InN lattice under hydrostatic pressure predict a pressure coefficient of around 1.1 cm$^{-1}$/GPa and 1.6 cm$^{-1}$/GPa for $M_L$ and $M_H$ respectively,[36] which is much lower than the measured hydrostatic pressure coefficient of the $A_1$(TO) phonon of 4.3 cm$^{-1}$/GPa.[24] Consequently, no significant deviation from the strain profile of the $A_1$(TO) mode is expected for the combinational modes $C_L$ and $C_H$. The expected values are indicated in Fig. 2b by the dashed lines, which reveal a good agreement with the experimental values. Furthermore, we note that the two-phonon density of states calculated by Reparaz *et al.* exhibits a pronounced double peak around 660 cm$^{-1}$,[24] which provides additional support for our interpretation of the $C_L$ and $C_H$ peaks as two-phonon combinational modes. The plasmon- and polariton-mediated localized excitation in TERS relaxes the far-field Raman selection rules and momentum conservation rule, which enables the observation of these modes exclusively in the near-field spectra of InN.

The explanation for the selectivity of the strong enhancement of the $A_1$(TO) and zone edge phonons $M_L$ and $M_H$ can be found in the nature of the scattering mechanism of TERS as discussed in a recent publication,[7] where the TERS process is described as a photon tunneling event. Due to the extent of the evanescent field created by the surface plasmon polaritons, the region of interaction is highly localized, as shown in Fig. 2a, well below the diffraction limit. Therefore, the scattering area is localized in a few-nanometer-sized region at the surface. This region includes the breaking of translational symmetry by the surface and the surface roughness and thus combination with K/M point modes. However, it should be noted, that the complex momentum carried by localized surface plasmons and surface plasmon polaritons and its conservation in the photon tunneling process is still a matter of scientific debate.[37,38,39]

Three additional weak Raman modes appear in the near-field spectra of the MEPA sample between 700 cm$^{-1}$ and 750 cm$^{-1}$ (Fig. 1a). The frequencies of these Raman modes as a function of the spatial position of the line scan are displayed together with the previously discussed modes $C_L$ and $C_H$ in Fig. 3 and compared to calculations of the two-phonon density of states (2PDOS) of InN in Ref.[24] Based on the reported calculations of the phonon dispersion relation, we tentatively assign the mode at about 700 cm$^{-1}$ to a second order combinational mode involving $E_2$(high) + $M_H$, with a Raman shift of 492 cm$^{-1}$ + 213 cm$^{-1}$ = 705 cm$^{-1}$. The other two peaks do not match any simple energy combinations of phonon modes but are located within the frequency range of the non-vanishing 2PDOS. In fact, the highest mode near 750 cm$^{-1}$ corresponds to the calculated high frequency onset of the 2PDOS in the right panel of Fig. 3. Consequently, we suggest that they also originate from second order processes.

| Phonon | Near-field (this work) MEPA sample [cm$^{-1}$] | Near-field (this work) MBE sample [cm$^{-1}$] | Far-field Ref.[24] [cm$^{-1}$] | Predicted Ref.[24]+Ref.[31] [cm$^{-1}$] |
|---|---|---|---|---|
| $A_1$(TO) | 452 | 452 | 451 | |
| $E_1$(TO) | 480 | | 478 | |
| $E_2$(high) | | 491 | 493 | |
| $C_L = A_1$(TO)+$M_L$ | 648 | 646 | | 644 |
| $C_H = A_1$(TO)+$M_H$ | 664 | 660 | | 663 |

**Tab. 1:** Phonon frequencies of InN. Near-field values are measured by TERS in the MEPA and MBE grown samples. Far-field and expected phonon frequencies based on literature values and phonon dispersion relation calculation as described in Eq. 1 and in the text.

Following the discussion of the individual modes, two important experimental observations of this work should be highlighted: First, the significant signal enhancement of TERS in InN under specific excitation conditions, *i.e.* bulk gold tip with 633 nm laser excitation, and second, the exclusive appearance of strong second order combinational modes in the near-field Raman spectra. In order to understand these observations, we consider the interaction between all constituents of the near-field scattering process as described in Ref.[7]. The complex scattering mechanism comprises of one particle and three quasiparticles: photons, phonons, localized surface plasmons (LSPs), and surface plasmon polaritons (SPPs). All these (quasi-)particles populate the tip-sample junction, making the scattering process particularly sensitive to the individual surface properties of the samples. Depending on the free surface charge carriers of the studied material, the plasmon population can interact weakly or strongly with the LSPs and SPPs of the tip apex, thus influencing the electron-phonon coupling and the amplification of the Raman signal.

The mechanism of near-field light-matter interaction, described in details in the Support Information of Ref.[7], considers a complex resonant condition composed by three energy levels: the exciting photon energy (i) and both the plasmon and polariton energies of tip (ii) and sample (iii). A resonant matching can occur when the plasma frequency of the sample (iii) is close to the frequency of the surface plasmon polaritons, which populate the tip apex (ii). In particular, the plasmonic energy states of the Hamiltonian $\mathcal{H}_{\text{pol-ion}}$, *i.e.* tip plasmons and plasmons at the InN surface, become resonant.[7] Due to the match of the energy states, the denominator of the tip-enhanced Raman scattering probability goes to zero.[7] Comparable to canonical resonant Raman scattering, where the resonance condition strongly increases the electron-phonon coupling, here, the plasmon-phonon coupling increases which relaxes the Raman selection rules and enables the observation of forbidden and weak modes. The simplest plasmon-resonant TERS configuration is the one composed by a gold substrate and a gold tip, which is known in TERS literature as "gap mode" configuration and it is usually adopted for studying 2D materials and single molecules positioned inside the tip-substrate gap.[40,41,42]

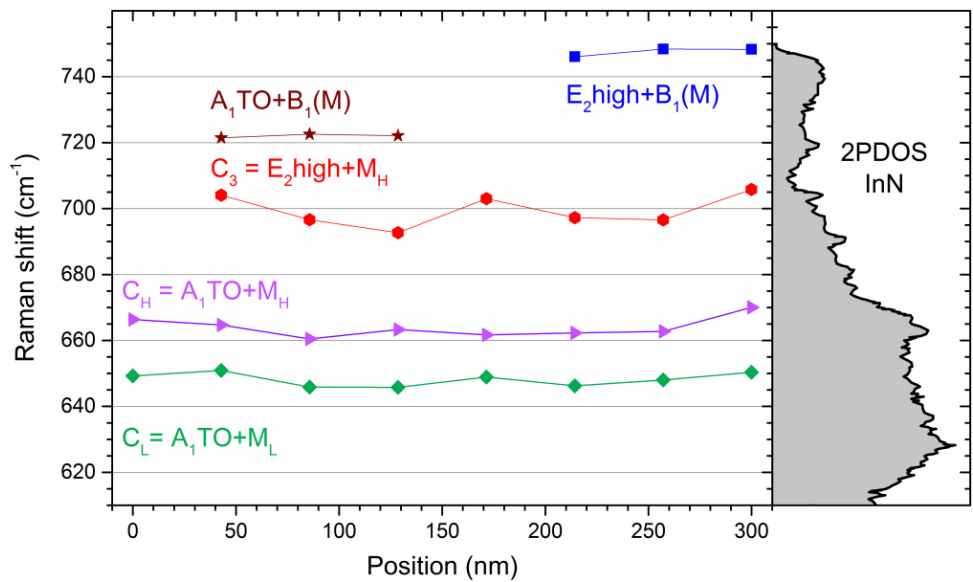

**Fig. 3** (left) Raman shift values for phonon modes in the high-frequency region of the MEPA sample as a function of the spatial position along the line scan. (right) Comparison with calculations of the two-phonon density of states (2PDOS) in Ref.[24] displayed in the same frequency range.

The particular case of TERS in InN discussed in this work likely presents a similar plasmon-resonant configuration due to the strong charge accumulation at the surface of InN.[43,44,45] It is important to note, that the observed enhancement can neither be explained by off-resonant near-field Raman scattering nor by resonant far-field Raman scattering alone. In order to underline this point, we have varied the excitation wavelength in both near-field and far-field configurations. When the gold tip is retracted from the InN surface, the signal enhancement disappears; the same applies, when the laser excitation is changed from 633 nm to 532 nm while the tip is engaged. This demonstrates that the strong near-field Raman scattering in InN is a resonance effect with plasmonic and polaritonic mediation that strongly benefits from the high surface charge carrier concentration in InN. Consequently, the enhancement factor of TERS strongly depends on the surface properties of the sample, which also explains the pronounced differences between the studied MEPA and MBE samples.

Following this discussion, we can classify the tip-sample interactions in TERS experiments in three different categories: "strong" plasmonic interaction regime, "weak" plasmonic interaction regime, and "non"-plasmonic interaction regime. We want to emphasize that with the simplified terminology of "plasmonic interaction", we also refer to the "polaritonic interaction" due to the simultaneous population of SPPs and LSPs in the tip-sample junction. An example of the strong plasmonic interaction regime is plasmon-resonant TERS in InN as studied in this work and also investigated in Ref.[21], where Raman peaks, not present in the far-field spectrum, appear in the TERS spectrum. The strong plasmonic interaction regime requires all three energy levels, i.e. laser excitation (i), plasmonic tip (ii), and plasmons of the materials (iii), to have a comparable energy. A schematic illustration of the energy levels of these three processes which are relevant for the plasmonic resonance condition in TERS is displayed in Fig. 4.

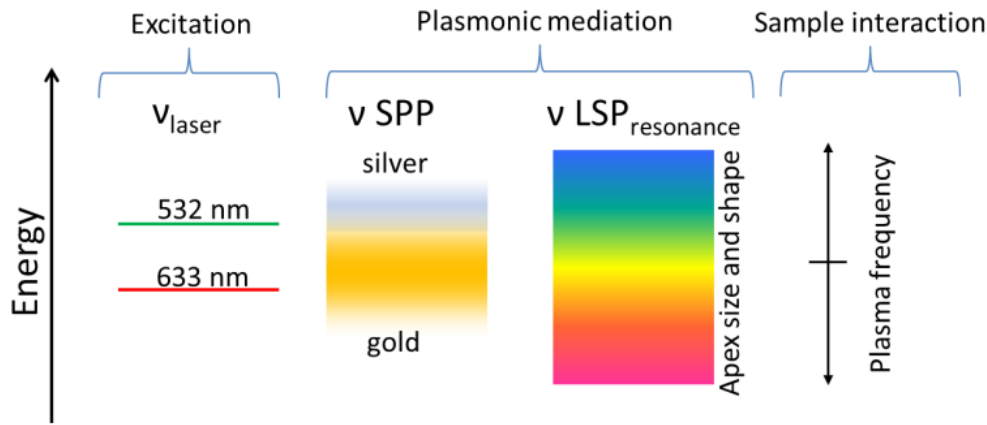

**Fig. 4** Schematic illustration of the energy levels involved in (plasmon-resonant) TERS for excitation (left), plasmonic mediation (center) and sample interaction (right). For the excitation energy, two common laser wavelengths are indicated as an example. The plasmonic and polaritonic energy levels depend on the tip material, apex size and shape. The plasma frequency depends on the sample's material and surface charge accumulations.

In this work, we achieve such strong plasmonic interaction by using a 633 nm excitation laser which is in resonance with the SPPs (ii) of gold,[46] and furthermore the LSPs of the tip apex (ii) depending on its geometrical shape[47], and the local SPPs (iii) of InN. Despite reports that the plasma frequency of InN spans from the mid-infrared to the near-infrared as a function of the carrier concentration,[48,49] the localized region of near-field interaction implies a probing depth of only a few nanometers below the surface. Thus, the bulk carrier concentration $n_{bulk}$ is not suitable to estimate the surface plasma frequency. As reported in Ref.[50], the charge carriers accumulate in the last 6 nm below the surface. This leads to an increase of the carrier concentration at the surface $n_{surface}$ of about one order of magnitude. Due to this surface accumulation layer (SA-layer), the material becomes a degenerate semiconductor at the surface with a free carrier concentration well above the Mott density within the interaction depth of TERS.[51] For the experimentally determined bulk carrier concentration in the MEPA grown InN sample of $n_{bulk} \approx 0.8 - 1 \times 10^{20}$ cm$^{-3}$, we thus estimate the magnitude of the SA-layer carrier concentration to be in the low $10^{21}$ cm$^{-3}$ regime over the TERS probing depth. Using the equation $\omega_p^{surface} = \sqrt{\frac{Ne^2}{\varepsilon_0 m_e(\varepsilon_\infty + 2\varepsilon_m)}}$,[52] with $\varepsilon_\infty = 7.84$ the high-frequency dielectric constant and $m_e = 0.07 m_0$ the effective electron mass,[53] we thus obtain SPP wavelengths between $1240 - 620$ nm for surface carrier concentrations between $n_e^{surface} = 0.5 - 2 \times 10^{21}$ cm$^{-3}$, respectively. This wavelength regime extends to the optical range and thus gives rise to the plasmon-resonant TERS.

For the MBE sample, the surface carrier concentration $n_e^{surface} \approx 1.0 - 1.4 \times 10^{19}$ $cm^{-3}$ is estimated from the phonon plasmon coupled mode at 430 cm$^{-1}$ (not shown) which corresponds to a spectral range for SPPs of $\omega_{SP} \approx$ 1100–1350 cm$^{-1}$ or 7400 – 8800 nm. This wavelength range is too far in the IR to obtain a plasmon resonance with the excitation laser. However, there are several aspects which can contribute to a local increase in carrier concentration, allowing for the observation of the PR-TERS effect also in the MBE sample. First, local fluctuations of the carrier concentration in combination with surface roughness result in a local modification of the shape factors in the plasma frequency equation. Second, the c-plane surface of InN is reported as degenerated semiconductor even for low bulk carrier concentrations. In particular, In-In dangling bonds lead to a Fermi-pinning above the conduction band minimum at 1.3-1.6 eV.[54] Furthermore, Ramsteiner et al. noticed in the case of degenerately-doped GaN that the Fermi energy is overestimated by a comparison with the plasmon coupled-mode spectra. They observed a considerable increase of the charge density in photo-current measurements even with below-bandgap excitation.[55] These observations indicate that the degeneracy steps in at lower $n_e$ than expected. We therefore conclude that the surface inhomogeneity, coupled with the carrier density increase under illumination, could account for a surface charge accumulation (SCA), which is sufficient to enable the plasmon-resonant TERS process even in InN with nominally lower free carrier concentration.

Finally, we briefly consider the cases of weak- and non-plasmonic interaction. A detailed study of an example of the weak plasmonic interaction regime has been reported in Ref.[7] for TERS on graphene and carbon nanotubes, where the intensity of the far-field Raman peaks are enhanced by a factor of two when approaching the tip. In that case, the three energy levels (i), (ii) and (iii) are nearly in resonance, but do not match completely since the plasma frequency of carbon allotropes lies in the IR.[56,57] TERS in ZnO, using e.g. a gold tip or a 633 nm excitation laser, can be chosen as an example for the non-plasmonic interaction regime, where the lack of enhancement in our own experiments underlines the missing near-field resonance condition. Only recently, TERS on two monolayers of ZnO has been reported in case of plasmonic resonance with local surface electronic states generated between the silver surface of the substrate and the 2D ZnO layers.[58] Using three different laser excitation energies and two plasmonic materials, Liu *et al.* were able to find a resonance with localized electronic interface states and observed tip-enhanced resonance Raman scattering (TERRS) of the ZnO bilayer.[58] The previous examples demonstrate that the classification as "strong", "weak", or "non-" plasmonic interaction is not an intrinsic material property, but rather depends on the correlation between laser excitation energy, tip material and dimensions, and plasma frequency of the sample.

**Conclusions**

In conclusion, we have shown that a strong near-field light-matter interaction occurs for tip-enhanced Raman scattering of InN, which is reproducible and independent of the growth technique of the samples. In the spectral range around 650 cm$^{-1}$, two distinct modes were observed in the near-field spectra, which exhibit particularly strong enhancement for plasmon-resonant excitation. The frequency of these modes matches a maximum in the two-phonon density of states of InN. We have identified these modes as two-phonon combinational modes $C_L$ and $C_H$, consisting of the optical phonon $A_1$(TO) and the acoustic phonons $M_L$ and $M_H$ near the M point of the Brillouin zone. The unusually strong enhancement of these modes was attributed to a resonance between the exciting photons with the plasmons and polaritons of the tip and the InN surface. Based on these observations we conclude that plasmon-resonant near-field Raman characterization of semiconductors with large surface charge accumulations is particularly powerful due to strong resonance enhancements and thus has tremendous potential for the characterization of In clusters and InN inclusions in technologically relevant InGaN heterostructures and nanostructures with high free carrier concentrations.

**Acknowledgments**

We thank Prof. N. Esser for fruitful discussions.